\documentclass[draft]{agujournal2019}
\usepackage{url} 
\usepackage{lineno}
\usepackage[inline]{trackchanges} 
\usepackage{soul}
\usepackage[version=3]{mhchem}
\usepackage{amsmath}
\usepackage{setspace}
\journalname{arxiv.org}

\begin{document}
	
	%
	%
	
	
	\title{Effect of 2009 major SSW event on the mesospheric CO$_2$ cooling}
	
	%
	%
	
	
	
	\authors{Akash Kumar\affil{1}, M V Sunil Krishna\affil{1,2}, Alok K Ranjan\affil{1,3}}
	
	
	\affiliation{1}{Department of Physics, Indian Institute of Technology Roorkee, Roorkee-247667, India}
	\affiliation{2}{Centre for Space Science and Technology, Indian Institute of Technology Roorkee, Roorkee-247667, India}
	\affiliation{3}{Space and Atmospheric Sciences Division, Physical Research Laboratory, Ahmedabad - 380009, India}

	
	
	
	\correspondingauthor{M V Sunil Krishna}{mv.sunilkrishna@ph.iitr.ac.in}

	\justify
	

	\begin{keypoints}
		\item The circulation changes during a major SSW event can cause an increase in mesospheric CO$_2$ density during peak warming period. 
		\item The CO$_2$ infrared radiative flux is found to be anti-correlated with CO$_2$ density during the main phase of SSW event. 
		\item The CO$_2$ 15 $\mu$m radiative cooling is dynamically influenced by the changes in temperature and atomic oxygen density in addition to the change in CO$_2$ abundance.
	\end{keypoints}
	
	%
	%
	
	%
	%
	
		
		\begin{abstract}
			
			Carbon dioxide (CO$_2$), an important trace species that is gradually increasing in the atmosphere due to anthropogenic activities, causes enhanced warming in the lower atmosphere. The increased concentration of CO$_2$ in the upper atmosphere results in enhanced radiative cooling rates leading to the contraction of the upper atmosphere. Due to its long lifetime and large vertical gradient, CO$_2$ concentration is also influenced by large dynamic events. We report a startling case of variability in CO$_2$ density and its infrared radiative cooling rates in the mesosphere and lower thermospher during a major sudden stratospheric warming (SSW) event. A counter-intuitive connection between CO$_2$ density and resulting CO$_2$ radiative cooling has been observed during the 2009 major SSW event. The behaviour of CO$_2$ cooling rates during such a dramatic events draw attention to our current understanding of CO$_2$ infrared cooling variation and its connection to changes in CO$_2$ concentration. The significance of temperature and atomic oxygen variability in the observed cooling patterns despite changes in CO$_2$ concentration, is also highlighted. 
			
		\end{abstract}

		\section*{Plain Language Summary}
		CO$_2$ is a greenhouse gas that is gradually increasing in the earth's atmosphere due to anthropogenic activities causing variations in the thermal structure of the atmospheric layers. The collision of CO$_2$ with other atmospheric constituents excites it in the vibrational state, which radiates the 15 $\mu$m band, therefore converting the kinetic energy into radiation and engendering cooling. Many dynamical events, such as sudden stratospheric warming events, influence the typical structure of the middle atmosphere ($\sim$30-110 km), including the transport of trace species such as CO$_2$. These events occur when vertically propagating planetary waves deposit momentum on breaking and lead to a change in circulation, primarily in the stratosphere. We have studied the changes that occurred in the CO$_2$ density and CO$_2$ infrared radiative cooling in the mesosphere and lower thermosphere at 60$^{\circ}$ - 70$^{\circ}$ N during the 2008-2009 winter. The CO$_2$ density shows variations in the polar mesosphere during the SSW event. These variations in CO$_2$ density are due to circulation changes, causing an ascent (descent) of CO$_2$ rich (poor) air-mass into higher altitudes during (after) these events. The effect of CO$_2$ variability on the CO$_2$ radiative cooling has also been analyzed during the major SSW event, and it is found that atomic oxygen, in addition to temperature variation, plays a crucial role in the CO$_2$ radiative cooling changes.

		%
		%

		\section{Introduction}
		Carbon dioxide (CO$_2$), a trace species, plays a crucial and important role in maintaining the thermal budget of the earth's atmosphere. Its increasing abundance in the lower atmosphere, due to anthropogenic activities, has resulted in global warming in the troposphere, thereby causing climate change \cite{lashof1990relative, solomon2009irreversible}. The increasing CO$_2$ abundance also reaches the upper atmosphere by means of advective transport and vertical eddy mixing \cite{chabrillat2002impact, yue2015increasing} resulting in a nearly constant volume mixing ratio (VMR) up to $\sim$80 km. The molecular diffusion and photolysis processes result in its rapid decrease above 80 km. Recent studies have confirmed the increasing CO$_2$ concentration in the atmosphere above the troposphere due to eddy diffusion/mixing \cite{garcia2016secular, lavstovivcka2023progress, qian2017carbon}. The increasing CO$_2$ concentration can significantly influence the compositional, thermal, and dynamical structure of the middle atmosphere (30-110 km) \cite{lopez2001non, wang2022climatology}. Due to its long lifetime (hundreds of days), CO$_2$ can also be influenced by the middle atmospheric circulation \cite{garcia2014distribution, lopez2000review} and acts as a dynamical tracer. The CO$_2$ enhancement in the upper atmosphere is due to upward circulation in the summer mesosphere, and reduced CO$_2$ is due to downward circulation in the winter mesosphere \cite{chabrillat2002impact, wang2022climatology}, thus it closely follows the mesospheric circulation.
		
		The role of CO$_2$ in cooling the middle atmosphere due to infrared radiative emission by its vibrational relaxation is now well understood \cite{castle2006vibrational, lopez2001non, mlynczak2010observations, mlynczak2022cooling, castle2012vibrational, roble1989will}. The cooling by CO$_2$ dominates in the middle atmosphere compared to cooling by other trace species such as ozone and water vapor \cite{kuhn1969infrared}. It is the fundamental band of CO$_2$, centered at 15 $\mu$m, which dominantly contributes to the radiative cooling compared to the other hot bands, primarily in the mesosphere and lower thermosphere (MLT) \cite{houghton1969absorption, dickinson1984infrared}. The transition from the vibrationally excited state of CO$_2$ corresponding to vibration-rotational mode ($\nu_2$) results in 15 $\mu$m emission, which leaves the optically thin upper atmosphere and eventually causes cooling. This radiative cooling by CO$_2$ contribute to the decrease of mean temperature resulting in the contraction of the upper atmosphere and therefore, leads to changes in the atmospheric structure \cite{akmaev2006impact, mlynczak2022cooling, roble1993greenhouse}. 
		
		The collisions of CO$_2$ molecule with the oxygen (O$_2$) and nitrogen (N$_2$) molecules, and also with the atomic oxygen (O) excite the CO$_2$ into vibrational state ($\nu_2$) resulting in the vibrational-translational energy transfer \cite{lopez2001non}. The vibration-rotation energy transfer rate constants through CO$_2$-O collisions are much larger than those for CO$_2$-O$_2$ and CO$_2$-N$_2$ collisions in the MLT \cite{lopez2001non, wintersteiner1992line}. The collision of CO$_2$ with O consequently, is the main process exciting CO$_2$ into its vibrational mode, $\nu_2$. The role of O in the excitation and quenching of the CO$_2$ vibrational mode $\nu_2$, and therefore in the radiative cooling, is very crucial \cite{dickinson1984infrared, mlynczak2000contemporary, sharma1990role, shved2003measurement}, and any variation in its abundance can influence the thermal structure of the MLT \cite{mlynczak2000contemporary, mlynczak2018updated, ward1993role}.

		It is evident from the earlier studies that the trace species are affected by transport, waves, tides, and other extreme dynamical events \cite{holton2000role, liu2002study, smith2012global, xu2003numerical}. Among these extreme dynamical events, the sudden stratospheric warming (SSW) events are noteworthy, and they impact the atmosphere over a wide scale, demonstrating coupling between different atmospheric layers \cite{chandran2014stratosphere, funke2010evidence, yiugit2016review}. The SSW events occur due to the interaction between stratospheric mean flow and vertically propagating planetary waves (PW) \cite{baldwin2021sudden, matsuno1971dynamical, liu2005dynamical, schoeberl1978stratospheric}. This interaction slows down or sometimes reverses the zonal-mean zonal wind producing adiabatic warming in the polar stratosphere and cooling in the mesosphere, consequently, affecting the mean structure of the atmosphere extensively \cite{gao2011temporal, kodera2016absorbing, kumar2024influence, scherhag1960stratospheric, siskind2005observations, wang2019winter, zulicke2013structure}. The widely accepted criteria defining a major SSW event include the poleward temperature increase and reversal of zonal-mean zonal winds at 60$^{\circ}$ latitude and at pressure level of 10 hPa \cite{butler2018optimizing, butler2015defining, charlton2007new}. During an SSW event, the reappearance of the polar stratopause at middle mesospheric heights, after descending and breaking in stratosphere, is known as a major SSW event with elevated stratopause (ES) \cite{limpasuvan2012roles, siskind2007recent}.
		
		In the present study, the variation in the trace species such as CO$_2$ and infrared cooling by CO$_2$ fundamental band emissions at 15 $\mu$m is investigated during the major SSW event that occurred in the 2008-2009 winter. As mentioned earlier, CO$_2$ is the main constituent for the mesospheric cooling, and any variation in its abundance can significantly impact the thermal structure of the MLT region. Therefore, the study of the variation in CO$_2$ and its associated 15 $\mu$m IR radiative cooling emissions (hereafter: CO$_2$ IR cooling) during these extreme dynamic events can help us understand the lower and upper atmospheric coupling and the energy budget of the MLT region more accurately. We present the first observational evidence of the CO$_2$ 15 $\mu$m IR radiative cooling change during a major SSW event and a peculiar case in which changes in the CO$_2$ concentration and CO$_2$ IR cooling rate are anti-correlated in the MLT during a major SSW event. The role of temperature and concentration changes is explored in understanding the anti-correlation observed.

		\section{Data and Methodology}
		To investigate the impact of the 2009 major SSW event on the energetics, dynamics, and transport of the mesospheric region we have utilized the observational data of the temperature, CO$_2$ IR cooling and O density derived from Sounding of the Atmosphere using Broadband Emission Radiometry (SABER) instrument on-board NASA's Thermosphere, Ionosphere, Mesosphere, Energetics and Dynamics (TIMED) satellite. The instrument's features are described in detail by \citeA{russell1999overview} and \citeA{esplin2023sounding}. The temperature data is retrieved from the CO$_2$ 15 $\mu$m emission and is available from nearly 15 km to 110 km, whereas, the O density data is available from 80 km to 100 km in the SABER database. The CO$_2$ cooling data available between 30-140 km in the level 2B database has been used, which is selected mainly for the fundamental band of the main isotope ($^{12}C^{16}O_2$), as it is the dominant source contributing to the radiative cooling in the MLT \cite{dickinson1984infrared}. In this study, we have employed the above-mentioned datasets in the north viewing mode of SABER during the 2008-2009 winter having a major SSW event. 
		
		The model data for winds, O$_2$ and CO$_2$ concentration, and CO$_2$ IR cooling have been obtained from the "specified-dynamics" (SD) version of the thermospheric and ionospheric extension of the Whole Atmosphere Community Climate Model (WACCM-X) \cite{liu2018development}. The top boundary of this model is 500-700 km above the earth's surface, and the upward extension extends up to that altitude with the same vertical resolution as the standard version of the model (WACCM) below 0.96 hPa, but with a top pressure of $4.1\times10^{-10}$ hPa. The vertical resolution has been enhanced to one-quarter scale height above 0.96 hPa. Ionospheric transport, neutral wind dynamo, and the estimation of ion/electron energetics and temperatures are all included in WACCM-X. The non-orographic gravity waves forcing parametrization suggested by \citeA{garcia2017modification}, has been used in the WACCM-X for better understanding of the MLT dynamics and transport. 
		
		The development, validation and other details about the WACCM-X version 2.0 are given in \citeA{liu2018development}. The residual mean meridional circulation \cite{andrews1987middle} has been derived using the meridional and vertical components from SD WACCM-X to understand the transport of the trace species in the MLT. All the three wind components with the above-mentioned parameters for the middle atmosphere, zonally averaged between 60$^{\circ}$ and 70$^{\circ}$ N, are calculated using daily mean model output data from SD WACCM-X history files, available at NCAR (National Center for Atmospheric Research) Climate Data Gateway. The history files comprise the SD WACCM-X simulation with nudging of Modern-Era Retrospective Analysis for Research and Applications, Version 2 (MERRA-2) data from the surface up to approximately 50 km.
		
		To understand the changes in the energetics of the MLT, due to CO$_2$ variability induced by altered dynamics and transport during the 2009 major SSW event, CO$_2$ mixing ratio has been utilised from Atmospheric Chemistry Experiment Fourier Transform Spectrometer (ACE-FTS) database. ACE-FTS has employed a high resolution Fourier transform spectrometer and two imaging detectors on-board the SCISAT-1 satellite. ACE-FTS uses solar occultation measurement technique to determine the volume mixing ratios of various chemical constituents of the earth's atmosphere in an altitude range between 10-100 km in a polar orbit. The details about the instrument can be found in \citeA{soucy2002ace}, and about CO$_2$ retrieval in \citeA{foucher2011carbon}. The CO$_2$ mixing ratio in parts per million (ppm) derived from the ACE-FTS Level 2 version 4.1/ 4.2 dataset in the polar region (60$^{\circ}$ - 70$^{\circ}$ N) has also been converted to density (cm$^{-3}$) as per the scheme described in \citeA{finlayson1999chemistry}. Zonally averaged daily variation of all the aforementioned physical quantities has been used to study the evolution of the middle atmospheric mean state during an SSW event. 
		
		The 2009 SSW event was a record-breaking warming event that was accompanied by the splitting of the cold polar vortex due to planetary wave 2 amplification and saturation. On 24 January, the zonal wind reversed at 60$^{\circ}$ N and 10 hPa level, demonstrating a major SSW event. As previously stated, this event was classified as a major SSW event with elevated stratopause (ES). The effects induced by this event in lower and upper atmosphere are well reported \cite{gao2011temporal, kodera2016absorbing, kumar2024influence, manney2009aura, shepherd2014stratospheric, randall2009nox}. The entire 2009 SSW event has been divided into three phases: (1) pre-SSW period (10-17 January 2009) when there were no major changes in the stratospheric temperature and winds, (2) peak warming or SSW main phase (18-27 January) when the stratospheric temperature increases and zonal wind reverses, and (3) SSW recovery phase (28 January – 20 February) when the stratospheric zonal wind and temperature recover.

\section{Results and discussion}
	\subsection{2009 SSW event and ACE-FTS measurements}
		Sudden stratospheric warming (SSW) events can induce significant changes in the middle atmosphere conditions by means of dynamical changes and the transport of trace species. CO$_2$, a trace species, is the most dominant cooling agent in the MLT region in comparison to other species, as mentioned earlier. The resulting changes in CO$_2$ abundance along with the temperature variations during a major SSW can induce significant variability in the overall radiative cooling of the MLT region. We have examined how the 2009 SSW affected the IR radiative cooling emissions caused by the CO$_2$ fundamental band and present first observational evidence of the CO$_2$ 15 $\mu$m IR radiative cooling change during a major SSW event. 
		
		The daily variation in geomagnetic indices, and zonally averaged temperature at north pole (at latitudes $>$ 80$^{\circ}$ N) and zonal-mean zonal wind (at 60$^{\circ}$ N) during the 2009 winter are shown in Figure 1. 
		\begin{figure}
			\centering
			\noindent\includegraphics[scale=0.6]{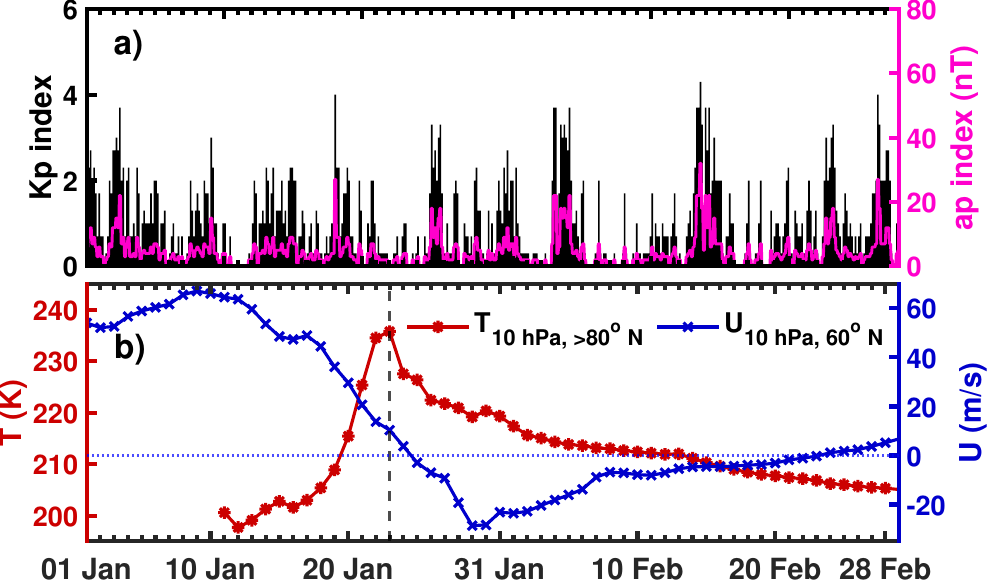}
			\caption{The geomagnetic indices, Kp (black) and ap (magenta) obtained from the OMNIWeb (a), and temperature (red starred line) and zonal-mean zonal wind (blue crossed line) at the defining latitude (60$^{\circ}$) and pressure level (10 hPa) derived from SABER and SD WACCM-X (b), respectively, during the January-February 2009. The vertical dashed line indicates the peak warming time over the polar stratosphere.}
		\end{figure}
		The geomagnetic conditions were nearly quiet as can be seen in the Figure 1(a). The variations in Kp and ap indices were low during the January-February 2009. Since most of the time Kp and ap indices were below 4 and 40 nT, respectively, the chosen time-frame for this study can be considered nearly geomagnetic quiet with only a few minor activities (19 and 26 January, and 4 and 14 February 2009). Although minor and short-term fluctuations in geomagnetic activity can affect the thermosphere and ionosphere at higher altitudes, variations in the mesosphere during the considered time period can be considered dominating by lower atmospheric changes. The sudden temperature increases above the polar cap (at latitudes $>$ 80$^{\circ}$ N) and reversal of the zonal-mean zonal wind over 60$^{\circ}$ N at 10 hPa level are characteristic of a major SSW, as depicted in Figure 1(b). During pre-SSW period the temperature at 10 hPa was around 200 K. The temperature has increased rapidly and peaked on 23 January, and zonal-mean zonal wind reversed at the time of peak warming. The polar stratospheric temperature gradually decreases and the zonal winds also recover slowly during the recovery phase of the SSW event.
		
		It is well established that during an SSW event, the temperature variation in the polar stratosphere is primarily driven by the breakdown of the zonal circulation, which in turn, results from the saturation of the planetary waves disturbing mean wintertime circulation \cite{limpasuvan2012roles, matsuno1971dynamical, schoeberl1978stratospheric}. The upward propagating planetary scale waves dissipating in the upper stratosphere and lower mesosphere (USLM) region induce a westward wave forcing, which reverses the zonal mean wind and causes changes in mean temperature gradient \cite{liu2002study}. The reversed circulation induces adiabatic warming in the polar stratosphere and cooling in the polar mesosphere. The circulation change in the middle atmosphere have a further influence on the lower and upper atmospheric structures and dynamics. These fluctuations are known to have a significant effect on the concentration and transport of trace species in the MLT \cite{bailey2014multi, baldwin2021sudden, manney2009aura}.
		
		The variations in the ACE-FTS measured CO$_2$ mixing ratio (ppm) in the MLT, along with the latitudinal coverage of ACE-FTS measurements, zonally averaged between 60$^{\circ}$ and 70$^{\circ}$ N during the January-February 2009, are shown in Figure 2.
		\begin{figure}
			\centering
			\noindent\includegraphics[scale=0.7]{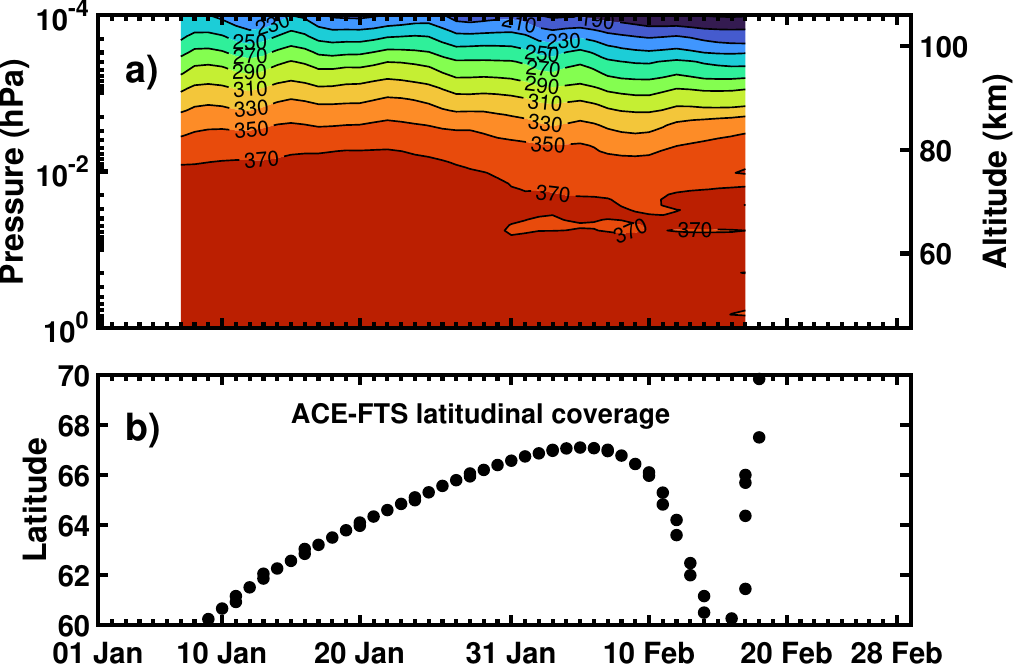}
			\caption{The ACE-FTS measured daily mean CO$_2$ mixing ratio (ppm) in the middle atmosphere (a), zonally averaged at the measurement latitudes between 60$^{\circ}$ and 70$^{\circ}$ N (b) during the January-February 2009.}
		\end{figure}
		During northern hemispheric SSW events, a model study indicated that the northern polar region reflects the most significant modifications in CO$_2$ concentrations in the MLT \cite{orsolini2022abrupt}. Therefore, CO$_2$ variability is investigated in the polar MLT using ACE-FTS data of CO$_2$ mixing ratio (ppm), available for the selected periods between 60$^{\circ}$ and 70$^{\circ}$ N. A significant decrease in the upper mesospheric CO$_2$ mixing ratio during the recovery phase after a small increase is seen during the 2009 SSW main phase compared to the pre-SSW values. The 370 ppm CO$_2$ mixing ratio level, for instance, was at $\sim$80 km, which rises above 80 km around 20 January and falls below 70 km after the end of January 2009. The next section delves deeper into anomalies associated with observed variations in CO$_2$ density (cm$^{-3}$) during the main phase.

		\subsection{Variations in CO$_2$ density and CO$_2$ IR cooling} 
		The anomalies in the ACE-FTS measured and SD WACCM-X estimated CO$_2$ density (cm$^{-3}$), and associated CO$_2$ IR cooling rates derived from 15 $\mu$m IR emission from the CO$_2$ fundamental band observed by SABER and the values estimated from SD WACCM-X, zonally averaged between 60$^{\circ}$ and 70$^{\circ}$ N during the 2009 SSW are shown in figure 3. 
		\begin{figure}
			\centering
			\noindent\includegraphics[scale=0.5]{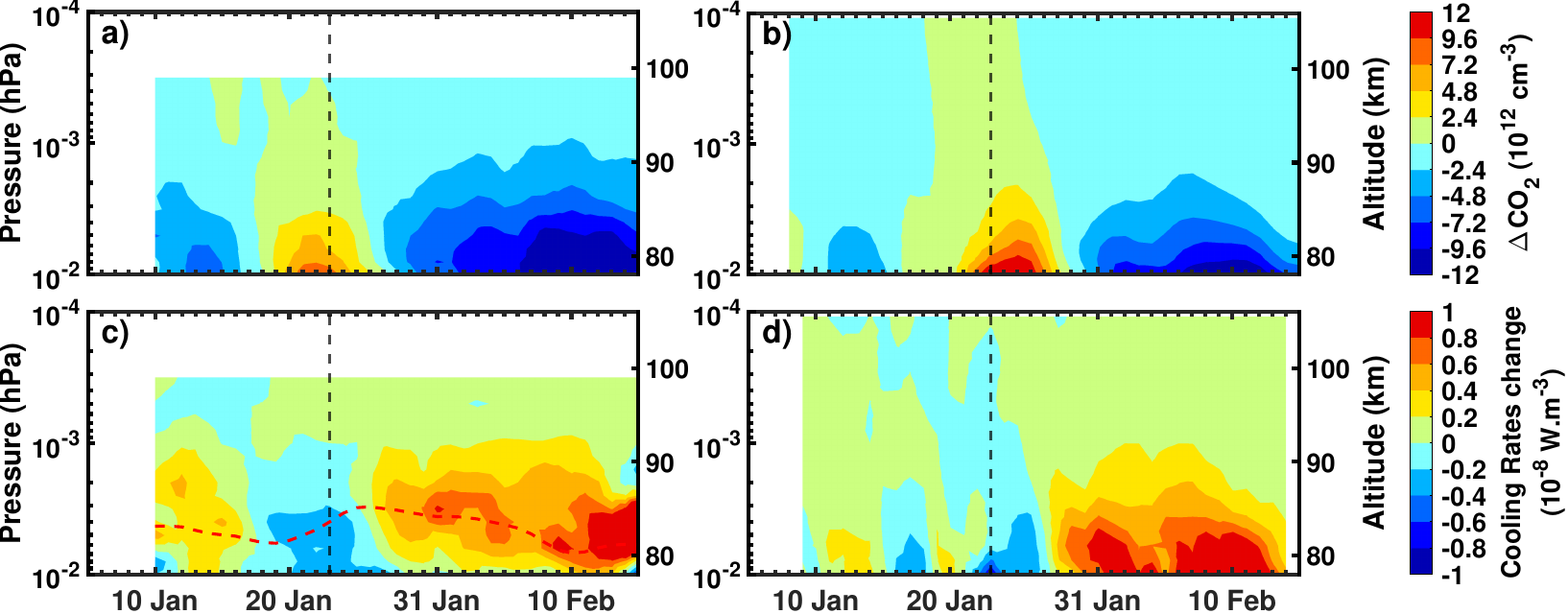}
			\caption{The anomalies in the CO$_2$ density (upper panels) measured by ACE-FTS (a), and calculated from SD WACCM-X (b), and IR cooling by CO$_2$ fundamental band at 15 $\mu$m (lower panels) observed by SABER (c), and calculated from SD WACCM-X (d), during the 2009 SSW. The red-dashed line indicate the altitude of peak CO$_2$ cooling. The vertical dashed line indicates the peak warming time over the polar stratosphere.}
		\end{figure} 
		The anomalies have been calculated by subtracting their daily mean values from averaged pre-SSW mean values (e.g., $ \Delta CO_2 = {CO_2 - \overline{CO_2}}$), where mean value ($\overline{CO_2}$) is the average of daily mean values during 10-17 January 2009 identified as pre-SSW period. The infrared cooling provided by the other bands of CO$_2$ did not show significant variations compared to that of fundamental band during the 2009 SSW period. Therefore, only infrared radiative cooling by 15 $\mu$m emission from the CO$_2$ fundamental band (CO$_2$ IR cooling) has been utilised for the present analysis. The CO$_2$ IR cooling anomalies have also been estimated following as per the method stated above.
		
		It can be seen that both the ACE-FTS measured (Figure 3(a)) and SD WACCM-X estimated CO$_2$ density (Figure 3(b)) increased during the peak warming period compared to the pre-SSW levels. However, the enhancement in SD WACCM-X estimated CO$_2$ density is found to be larger ($\sim2$ times) in comparison to that derived from ACE-FTS. During the recovery phase, a significant reduction in the ACE-FTS measured and SD WACCM-X estimated CO$_2$ density has also been seen; however, the lower values of decreased CO$_2$ density were estimated by SD WACCM-X. The ACE-FTS measured CO$_2$ density decrease was prolonged, whereas SD WACCM-X estimated decrease in CO$_2$ density was confined upto early days of February. Similarly, a depletion in the CO$_2$ IR cooling during peak warming period and enhancement in the recovery phase has been seen in the SABER observed (Figure 3(c)) and SD WACCM-X estimated CO$_2$ IR cooling rates (Figure 3(d)). Here also, SD WACCM-X estimated a smaller decrease in CO$_2$ IR cooling and a larger enhancement in CO$_2$ IR cooling than those observed by SABER during the peak warming and recovery periods, respectively. The altitude of peak CO$_2$ cooling was found to occur around 0.003 hPa ($\sim85$ km) in SABER CO$_2$ IR cooling observations, whereas it was around 0.01 hPa level ($\sim80$ km) in WACCM-X estimated CO$_2$ IR cooling profile (Figure 3(d)). The peak CO$_2$ IR cooling altitude ascended $\sim$2-3 km with decreased cooling rates during peak warming period and descended $\sim$2-3 km with increased cooling rates during the recovery phase from its pre-SSW levels.

		The most important and peculiar aspect of the 2009 SSW-induced effects on CO$_2$ and IR cooling is the anti-correlation between CO$_2$ density and the magnitude of the net CO$_2$ IR cooling rate, as presented in Figure 4(a).
		\begin{figure}
			\centering
			\noindent\includegraphics[scale=0.55]{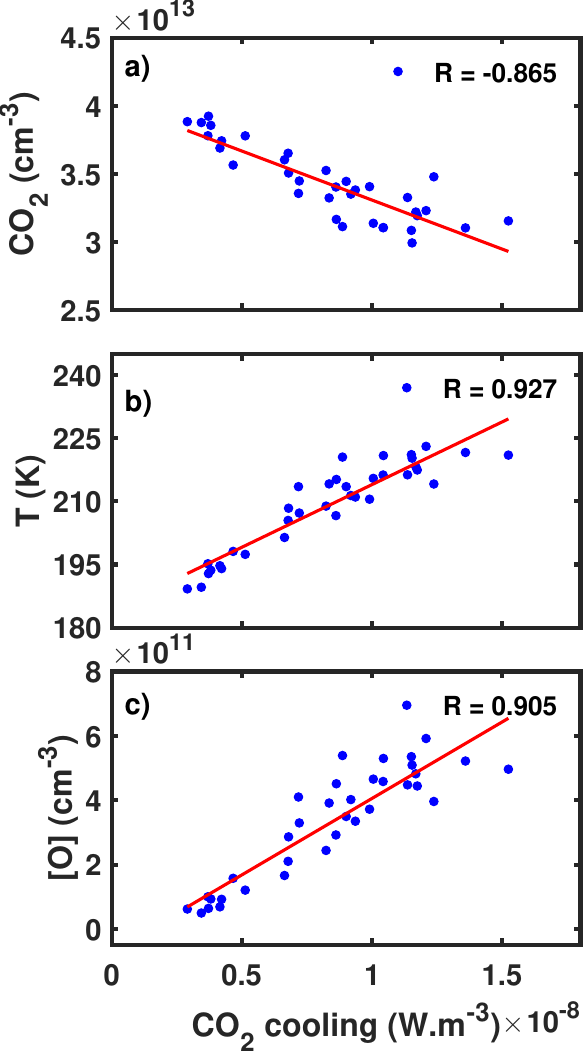}
			\caption{The correlation between-SABER derived daily mean CO$_2$ IR cooling and ACE-FTS-derived CO$_2$ density (a), SABER-derived CO$_2$ IR cooling and temperature (b), and CO$_2$ IR cooling and O density (c), zonally averaged between 60$^{\circ}$-70$^{\circ}$ N at 0.003 hPa level, during the 2009 major SSW.}
		\end{figure}
		The SABER derived CO$_2$ IR cooling, O density and temperature, and ACE-FTS derived CO$_2$ density have been utilised to compute the correlation between these quantities around the altitude of peak CO$_2$ IR cooling, at 0.003 hPa level ($\sim85$ km). The CO$_2$ IR cooling is strongly anti-correlated to the CO$_2$ density during the entire SSW period, with a correlation coefficient of -0.865 (Figure 4(a)). As expected, the temperature and O density were also highly correlated to the CO$_2$ IR cooling (Figure 4(b,c)) in the polar mesosphere during the entire period of study, with a correlation coefficient of 0.927 and 0.905, respectively. The correlation indicates that the changes in CO$_2$ density, temperature and O density are the possible contributor to the change in CO$_2$ IR cooling. Earlier studies suggested that gradually increasing CO$_2$ in the upper atmosphere leads to more cooling and contraction of the upper atmosphere \cite{mlynczak2022cooling}. However, the sudden increased/decreased CO$_2$ density during main/recovery phase of the 2009 SSW is found to be associated with the reduced/enhanced CO$_2$ IR cooling (Figure 3). Similar correlations have also been seen in the SD WACCM-X estimated CO$_2$ IR cooling, CO$_2$ and O density, and temperature (not shown here), despite different values of these quantities in comparison to their observed values.
		
		The large enhancement in CO$_2$ IR cooling despite reduced CO$_2$ density during the recovery phase indicates a crucial role of the changes in temperature and O density. Since CO$_2$ IR emissions strongly depend on the temperature, the anti-correlation between CO$_2$ density and CO$_2$ IR cooling also indicate the bearing of CO$_2$ density change on CO$_2$ IR cooling rates during the SSW event. Therefore, the contribution of individual parameters (e.g., temperature, O and CO$_2$) has been investigated to understand the observed variations in CO$_2$ IR cooling during the 2009 SSW event. Consequently, we can imply that the changes in internal forcing in the form of the kinetic temperature and composition perturbations, driven by the SSW events, could be responsible for the variability in observed CO$_2$ IR cooling. The effect underlined in the above paragraph can be categorized as short term variability in comparison to those suggesting an increase in the mesospheric cooling (reduced temperature) due to the slowly increasing CO$_2$ concentration over the past decades. It is important to know how the SSW event can influence the key parameters (T, O, CO$_2$, etc.) so as to conclude the role of each of these parameters in the observed CO$_2$ IR cooling. 
		
		\subsection{Circulation change affecting neutral density}
		The upper mesospheric concentration of CO$_2$ is lower in the winter polar region in general when compared to the summer hemisphere \cite{chabrillat2002impact} following the circulation, as mentioned earlier. The observed dramatic increase and decrease in the CO$_2$ density over and above the climatological values during the initial and recovery phases of the 2009 SSW event due to changes in the residual mean meridional circulation have been investigated and shown in Figure 5. 
		\begin{figure}
			\centering
			\noindent\includegraphics[scale=0.55]{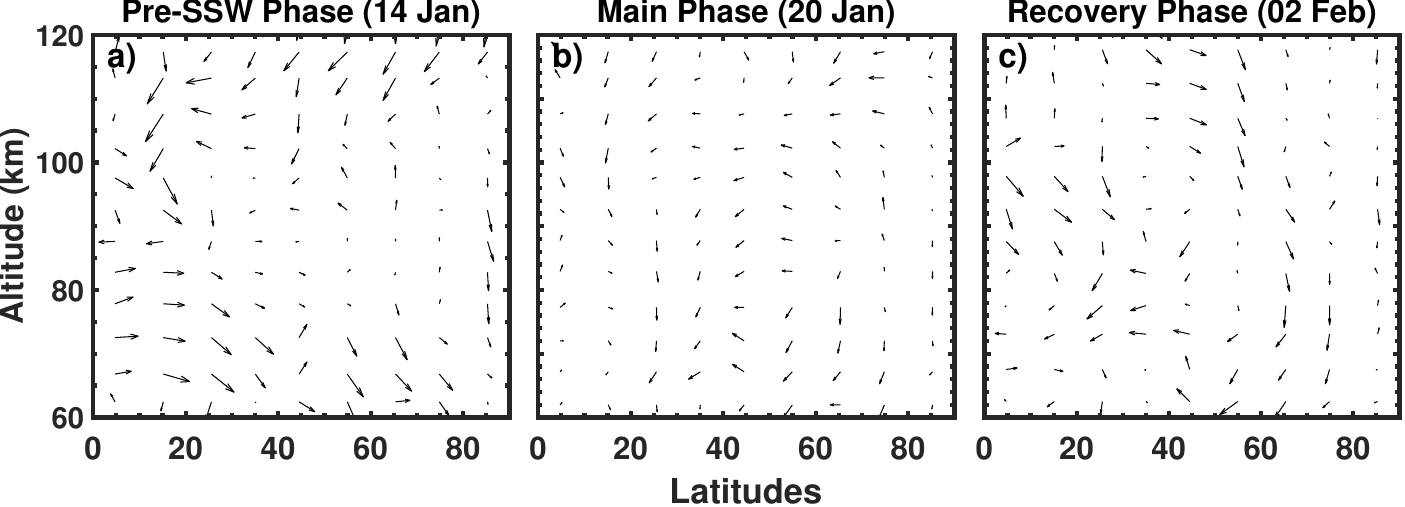}
			\caption{Wintertime residual mean meridional circulation derived from SD WACCM-X, before SSW onset (a), during the peak warming period (b), and in the recovery phase (c) of the 2009 SSW event.}
		\end{figure}
		Before the SSW period, the poleward/downward wind in the polar region represents the characteristic winter-time circulation (Figure 5(a)), which starts to reverse and becomes upward/equatorward in the polar MLT during the peak warming period (Figure 5(b)). During the recovery phase, the residual mean meridional circulation again becomes poleward/downward in the polar MLT with increased intensity (Figure 5(c)). It can also be noted that the intense downward mean meridional circulation extended from lower thermosphere into the upper stratosphere after the SSW event.
		
		The changes in the zonal mean winds are caused by a westward wave forcing that is generated due to the wave-mean flow interaction during SSW events. The westward forcing also affect the gravity-wave filtering, permitting more eastward gravity waves to reach into the MLT. This eastward wave forcing resulted in changes in the mean meridional circulation from poleward/downward to equatorward/upward during the initial phase of an SSW event \cite{liu2002study}. The westward planetary wave forcing gradually dominates the eastward gravity wave forcing in the MLT, which cause the mean circulation to become more poleward/downward and reformation of the elevated stratopause \cite{limpasuvan2016composite}. As mentioned earlier, CO$_2$ can also be influenced by the dynamics; these changes in the meridional circulation can cause the observed CO$_2$ variation in the MLT region during the SSW event.

		The observed changes in CO$_2$ density along with the temperature variation can be extended to understand the variability of other trace species, which are crucial for CO$_2$ IR cooling, at the peak CO$_2$ IR cooling altitude during the SSW period. The densities of CO$_2$, O$_2$, and O, with the temperature and CO$_2$ radiative flux calculated from the CO$_2$ IR cooling rates at 0.003 hPa pressure level ($\sim$85 km) also vary in a similar way, i.e., decreasing during SSW main phase, and increasing during the recovery phase (Figure 6). 
		\begin{figure}
			\centering
			\noindent\includegraphics[scale=0.6]{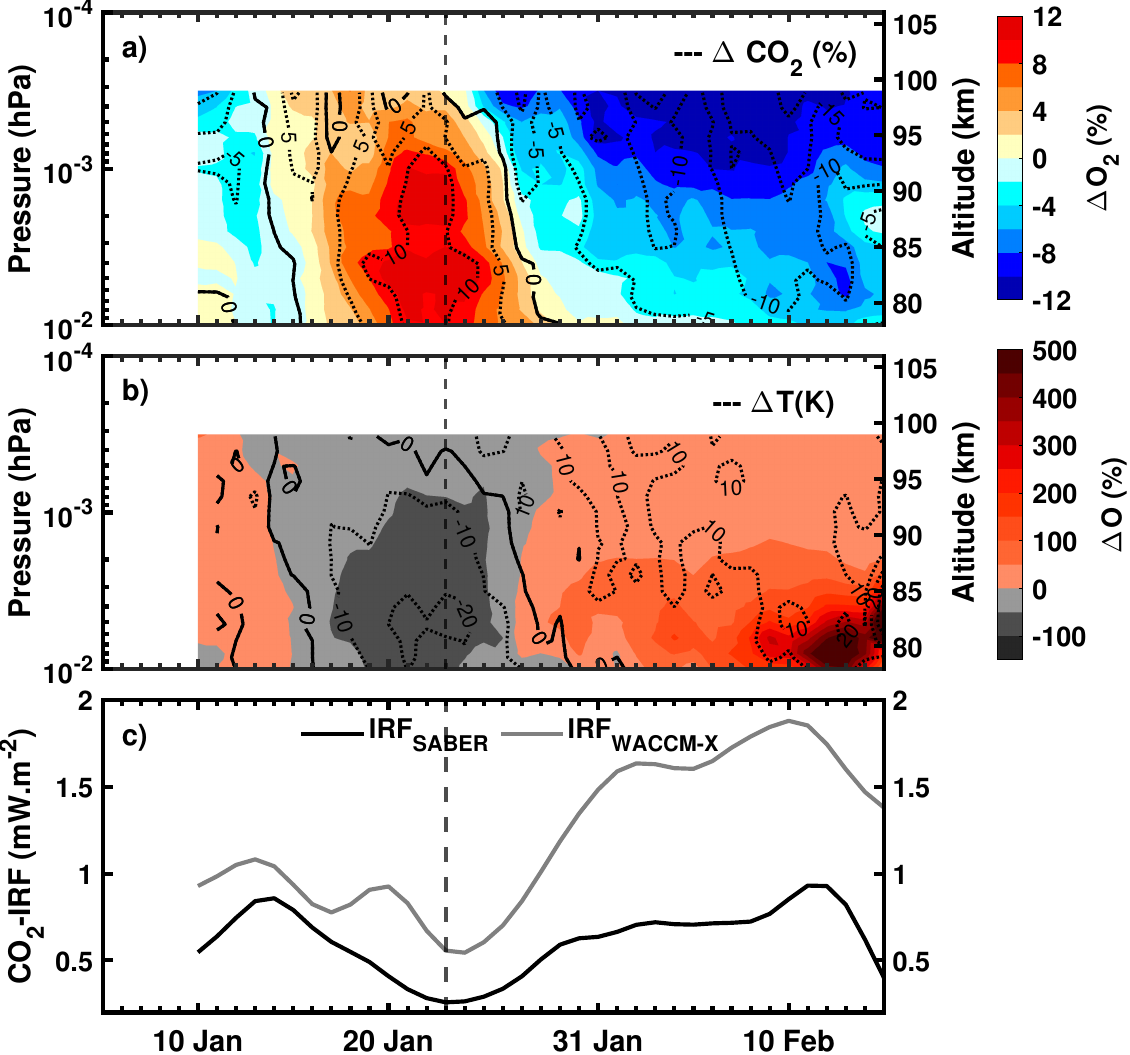}
			\caption{Variations in the ACE-FTS derived CO$_2$ density (contour lines) and SD WACCM-X estimated O$_2$ (colormap) density (a), SABER derived temperature (contour lines) and O density (colormap) (b), and CO$_2$ radiative flux (c) using SABER measurements (black line) and WACCM-X output (gray line) during the 2009 SSW. The vertical dashed line indicates the peak warming time over the polar stratosphere.}
		\end{figure}
		ACE-FTS derived CO$_2$ density shows $\sim$10-12$\%$ enhancement and $\sim$15$\%$ reduction during the SSW main and recovery phases from its pre-SSW level, respectively (contour lines, Figure 6(a)). SD WACCM-X estimated O$_2$ density also indicates $\sim$15$\%$ increase and $\sim$15-20$\%$ decrease during the SSW initial and recovery phases from its pre-SSW values, respectively (colormap, Figure 6(a)). It can also be noted that the enhancement in CO$_2$ and O$_2$ densities were large at lower altitudes, whereas, reduction were large at higher altitudes in the MLT.
		
		The upper mesospheric temperature has been decreased by $\sim$20 K during the peak warming period compared to the pre-SSW level. At the time of the ES formation at the mesospheric altitudes, i.e., during the recovery phase, the upper mesospheric temperature was enhanced by $\sim$20 K (contour lines, Figure 6(b)) in comparison to its mean values before the SSW event. A substantial decrease upto $\sim$~97$\%$, and an enormous enhancement upto $\sim$500$\%$ in SABER derived O density has been observed during the peak warming period and recovery phase, respectively, in comparison to the pre-SSW mean values (colormap, Figure 6(b). The maximum variations in the SABER derived temperature and O density are found at lower altitudes in the MLT. Compared to their pre-SSW mean levels, SABER derived CO$_2$ infrared radiative flux (IRF) was determined to be reduced by $\sim$20$\%$ during the SSW period and enhanced by $\sim$120$\%$ during the recovery phase. The SD WACCM-X derived CO$_2$ IRF was reduced by $\sim$50$\%$ and enhanced by $\sim$90$\%$ during the SSW main and recovery phases, respectively, regardless of the minor variations in the geomagnetic activity. Although, SD WACCM-X derived CO$_2$ IRF values were larger compared to those of SABER, the variations in CO$_2$ IRF values calculated using SD WACCM-X output were greater than those calculated using SABER observation during the entire examined period. SD WACCM-X estimated IRF was rapidly enhanced during recovery phase as compared to the SABER calculated flux. 
		
		As was previously stated, the variation in the trace species and temperature can be associated with the circulation change in the MLT during the SSW period. Before the SSW onset, the general winter-time circulation (Figure 5) along with the molecular diffusion causes the lower CO$_2$ mixing ratio above 80 km. During the peak warming period, upward/equatorward circulation (upwelling) pushes the CO$_2$ and O$_2$ rich air-mass to the higher altitudes in the polar region, resulting in the observed increase in CO$_2$ and O$_2$ densities during the SSW main phase, which remained at the same level for several days. This upward/equatorward circulation reverses back to downward/poleward (downwelling), bringing CO$_2$ and O$_2$ poor air-mass to the lower altitudes, resulting in the reduced CO$_2$ and O$_2$ densities in the polar MLT during the recovery phase of the 2009 SSW event. Similarly, a decrease and enhancement in the O density and temperature in the MLT during and after the SSW event were also associated with the upwelling and downwelling, respectively. The downwelling can transport O from its thermospheric reservoir into the lower mesosphere. The results showing variability of CO$_2$, O$_2$, and O during the 2009 SSW event are in agreement with the previous studies, which suggest the vertical transport of the trace species, e.g., NO$_x$, O, CO, ozone, etc., during and after the SSW events mainly due to vertical transport \cite{bailey2014multi, liu2002study, manney2009aura, orsolini2022abrupt, shepherd2014stratospheric, smith2011waccm}. The CO$_2$ IRF variability also reflects the above-mentioned changes in the thermal and compositional structure of the MLT. To understand the variability in CO$_2$ IRF (i.e., IR cooling rates), the contribution of changes in temperature and trace species that are critical for CO$_2$ IR cooling has been further investigated.
		
		\subsection{Change in CO$_2$ population ratio}
		Having established the defining features of SSW and the reason for the observed variation in the neutral composition, this study investigates the understanding of the cause for the observed changes in the CO$_2$ IR cooling/radiative flux during a major SSW event. As previously stated, and based on the molecular chemistry of CO$_2$, it is well known that O$_2$/N$_2$ and O play a critical role in the collisional excitation/quenching of CO$_2$ into its vibrationally excited state. Therefore, the ratio of the density of O and O$_2$ to the CO$_2$ density can help us understand the influence of composition change on the CO$_2$ IR cooling.  The SABER observed O density in combination with ACE-FTS derived CO$_2$ density has been used to calculate O/CO$_2$. Similarly, the ratio O$_2$/CO$_2$ has been calculated using the WACCM-X estimated O$_2$ density and the CO$_2$ density calculated from ACE-FTS observations. Hence, the variation in density ratios, O/CO$_2$ and O$_2$/CO$_2$, has been investigated with the changes in population ratio, CO$_2(01^10)$ to CO$_2(00^00)$, due to collisional excitation by O and O$_2$ (chemical reactions; R1 $\&$ R2), respectively, to comprehend the observed variability of CO$_2$ density and the CO$_2$ IR cooling.
		\begin{center}
			${ \ce{CO_2}(v_2=0) + \ce{M}(\ce{O_2}, \ce{N_2}) \hspace{0.4cm}
				\xrightarrow{k1} 
				\hspace{0.4cm}  \ce{CO_2}(v_2=1) + \ce{M}(\ce{O_2}, \ce{N_2}) }$ 
			\hspace{0.8cm} (R1)          
		\end{center}
		\begin{center}
			${ \ce{CO_2}(v_2=0) + \ce{O(^3P)} \hspace{0.9cm}
				\xrightarrow{k2} 
				\hspace{0.9cm}  \ce{CO_2}(v_2=1) + \ce{O(^3P)} }$ 
			\hspace{1cm} (R2)          
		\end{center}
		where
		$k1 = 9.6\times10^{-16} exp(-8.08\times10^{-4}T + 1.85\times10^{-2}T^2) $ and	$k2 = 3\times10^{-13}(T/300)^{1/2}$ \cite{lopez2001non}, are excitation rate coefficients (cm$^{-3}$s$^{-1}$) of the CO$_2$ molecule by the corresponding atoms/ molecules.
		
		Figure 7 illustrates the zonally averaged daily variation in the density ratios, O/CO$_2$ and O$_2$/CO$_2$ (left column), and the population ratio, CO$_2(01^10)$ / CO$_2(00^00)$, due to collisional excitation of CO$_2$ by O and O$_2$ (right column), respectively.
		\begin{figure}
			\centering
			\noindent\includegraphics[scale=0.5]{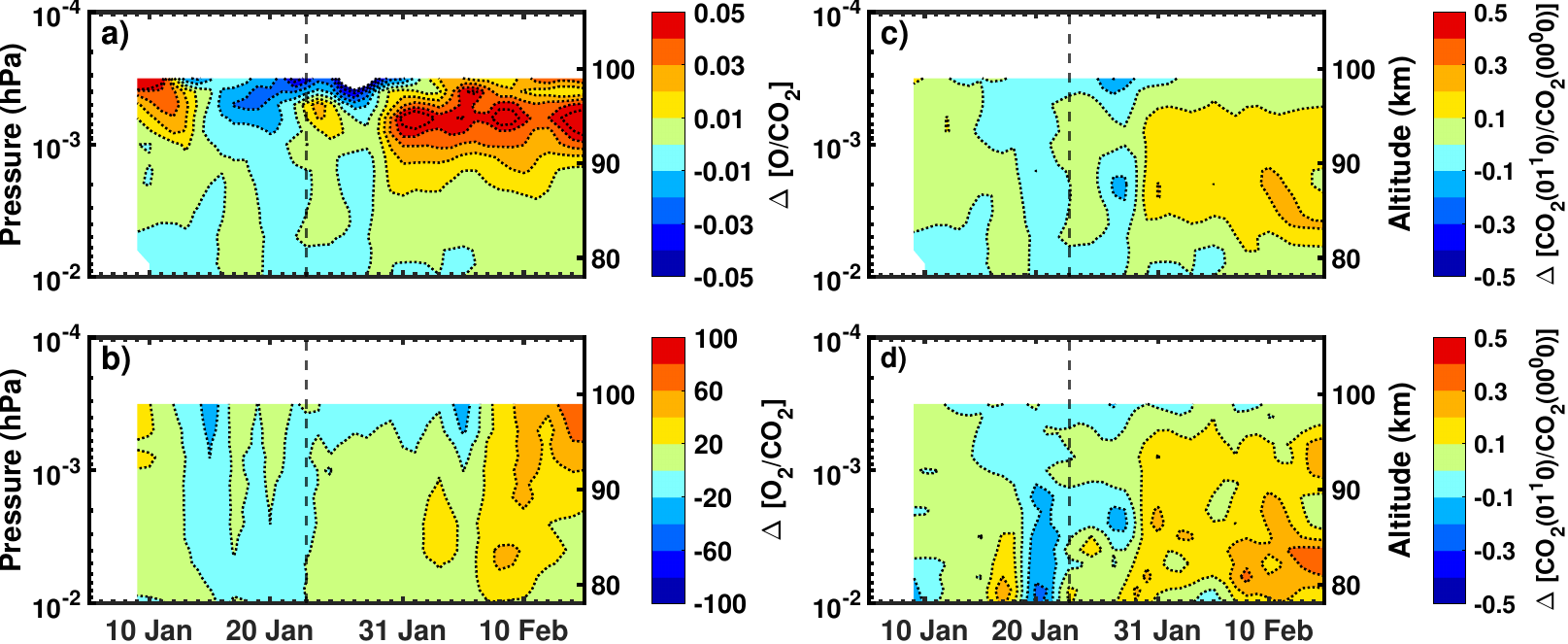}
			\caption{The anomalies in the O/CO$_2$ density ratio (a), O$_2$/CO$_2$ density ratio (b), population ratio due to collisional excitation by O (c), and O$_2$ (d), respectively, between 60$^{\circ}$-70$^{\circ}$ N, during the 2009 SSW event. The vertical dashed line indicates the peak warming time over the polar stratosphere.}
		\end{figure}
		It is clear from Figure 7 that the density ratios of O/CO$_2$ and O$_2$/CO$_2$ have been reduced during the peak warming period and enhanced in the recovery phase compared to their pre-SSW values. Although the larger variation in the O/CO$_2$ density ratio was above 0.001 hPa level ($\sim$ 90 km), at peak O altitude, the variation in O$_2$/CO$_2$ was also large below 0.001 hPa level in comparison to the O/CO$_2$ variation. Similarly, the population ratio due to collisional excitation by O and O$_2$ was also decreased during SSW main phase and increased in the recovery phase compared to their pre-SSW values. The maximum variation in the population ratio was between 0.01 and 0.001 hPa (80-90 km).
		
		The O density peak is also located near 97 km in the MLT. As stated earlier, during the SSW initial and recovery phases, the mean meridional circulation change causes O density in the MLT to decrease and increase, respectively. The variation in O density is much larger than that of CO$_2$ density (Figure 6), and indicated a decrease/enhancement in O/CO$_2$ ratio (Figure 7(a)) during initial/recovery phase of the SSW event. The density of O$_2$ and CO$_2$ decreases with altitude, therefore, an equatorward/upward meridional circulation increases and an intense poleward/downward meridional circulation reduces their density in the MLT during initial and recovery phases of the SSW event, respectively. The enhancement/reduction in O$_2$ density was slightly (5-7 $\%$) larger than that of CO$_2$ density (Figure 6(a)). Therefore, O$_2$/CO$_2$ ratio shows a net depletion/enhancement (Figure 7(b)) during initial/recovery phase of the SSW event. Consequently, the variation in the density ratio of O/CO$_2$ and O$_2$/CO$_2$, despite the CO$_2$ density changes during the peak warming and recovery periods, could be affected by the variation in the O and O$_2$ densities, which is also reflected in the CO$_2$ population ratio variations.
		
		The population ratio depends on the density of the species participating in the collisional excitation of CO$_2$ (here O and O$_2$/N$_2$), and on the temperature (included in $k_1$ and $k_2$). As a result, it can be noted that despite the enhanced CO$_2$ density, the reduced temperature and O density in the MLT (Figure 6) lead to the observed depletion in the population ratio during the peak warming period (Figure 7). Hence, we can infer that the concurrent decrease in mesospheric temperature and composition change during the peak warming period resulted in the lowering of the collisional excitation of CO$_2$ into the higher vibrational state, and leading to a low CO$_2(01^10)$ to CO$_2(00^00)$ ratio. Although reduced mesospheric temperature has been anticipated to be the primary contributor to decreased CO$_2$ IR cooling rates during the initial phase of the SSW event, the contribution of composition change in collisional excitation/quenching rates cannot be overlooked and needs to be explored.
		
		The aforementioned discussion pertains to the reduced CO$_2$ IR cooling in the mesosphere during the peak warming period of the SSW event, despite a small increase in CO$_2$ density. During the recovery phase, the increased mesospheric temperature at the time of ES reformation, and composition change due to downwelling resulted in increased collisional excitation of the CO$_2$ in its higher vibrational states and hence a larger CO$_2(01^10)$ to CO$_2(00^00)$ population ratio, which ultimately leads to enhanced CO$_2$ IR cooling in the MLT. Here also, the role of the large O density changes appears to support the enhanced temperature-induced variation in CO$_2$ IR cooling variability. Therefore, to investigate the relative contribution of the temperature and composition change, we analysed the population ratio (PR) for different conditions at the altitude of maximum CO$_2$ IR cooling variation (at 0.003 hPa, $\sim$85 km), as shown in Figure 8. 
		\begin{figure}
			\centering
			\noindent\includegraphics[scale=0.5]{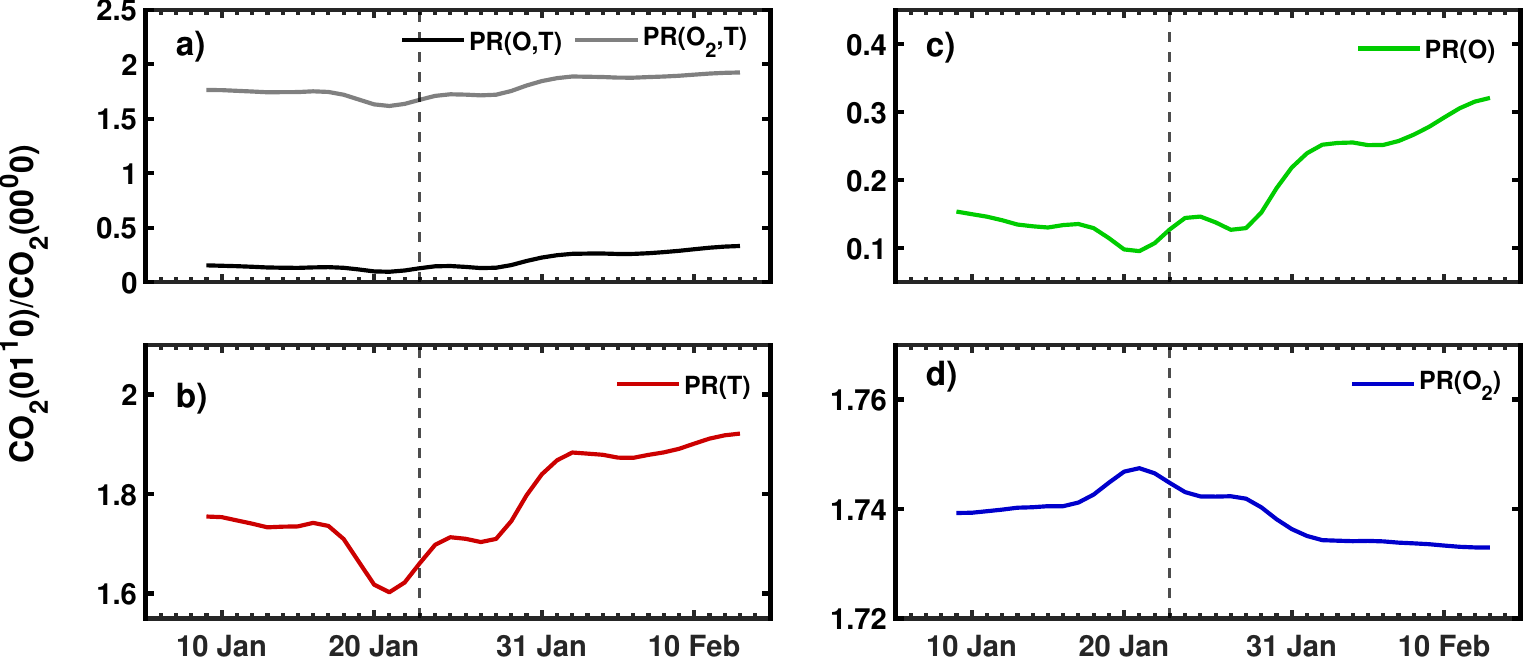}
			\caption{Variation in CO$_2(01^10)$ to CO$_2(00^00)$ population ratio due to temperature, O and O$_2$ change (a), due to only temperature change (b), due to only O density change (c), and due to only O$_2$ density change (d), respectively, at 0.003 hPa ($\sim$85 km) and 60$^{\circ}$-70$^{\circ}$ N, during the 2009 SSW event. The vertical dashed line indicates the peak warming time over the polar stratosphere.}
		\end{figure}
		By employing the pre-SSW mean values (averaged between 10 and 17 January) of the other parameters, the relative contribution in the PR caused by variations in one parameter, such as temperature, O, and O$_2$, has been evaluated separately. 
		
		The population ratio (PR) due to collisional excitation by O$_2$ in comparison to the collisional excitation by O, in combination with temperature, dominates below 0.001 hPa, as depicted in Figure 8(a). The variation in the PR due to only temperature change (Figure 8(b)), and due to only O density change (Figure 8(c)) follow the CO$_2$ IR cooling and CO$_2$ IRF pattern, i.e., depletion during the peak warming period and enhancement during the recovery phase. There is an opposite, although minor variation in the PR due to only O$_2$ density change (Figure 8(d)), which in combination with temperature variation results in depletion and enhancement in PR during peak warming and recovery periods, respectively. As seen in the figure 8(b,c), changes in the temperature and O density in the polar MLT are the main contributors in the PR variability, hence in the observed CO$_2$ IR cooling (see Figure 3).
		
		The CO$_2$-O$_2$ collisional excitation dominates the CO$_2(01^10)$ to CO$_2(00^00)$ population ratio below 95 km, as discussed earlier (Figure 8(a)). However, major depletion in PR during the peak warming period, compared to the pre-SSW level, is mainly affected by reduced temperature, followed by decreased O density. The changes in the PR due to O$_2$ variability, despite its large amplitude, were insignificant during the peak warming period as the overall PR decreased. Similarly, an enhancement in the PR during the recovery phase compared to its pre-SSW values is primarily driven by enhanced temperature. A small contribution to the PR by increased O density was seen that was enhanced during the recovery phase, whereas the impact of O$_2$ variation in enhanced PR was insignificant during the recovery phase. Despite the larger variation in PR caused by the O density changes following the SW event, its net effect was smaller than that caused by the temperature change in the MLT.
		
		Therefore, it is evident from the above discussion that, along with the temperature change and minor variation in the CO$_2$ density, the availability of atomic oxygen cannot be neglected in the overall CO$_2$ radiative emission processes. Reduced temperature during the initial phase of the SSW event, combined with the decreased O density, contributed to the reduced CO$_2$ IR cooling despite increased CO$_2$ density, and vice versa during the recovery phase, leading to the observed anti-correlation between CO$_2$ density and CO$_2$ IR cooling. Hence, the observed CO$_2$ IR cooling is dynamically controlled primarily by temperature changes, followed by atomic oxygen variations in addition to the changing density of the primary cooling agent, i.e., CO$_2$.

		\section{Summary}
		The rapid increase of CO$_2$ in the earth's atmosphere is the major controlling factor in the thermal budget of the whole atmosphere. Earlier studies reported that the long-term increase in the CO$_2$ concentration in the upper atmosphere leads to more radiative cooling, resulting in the contraction of the upper atmosphere. The present study exhibits results indicating that an SSW, an abrupt and dynamic event, has a significant impact on mesospheric CO$_2$ and resulting infrared radiative cooling.
		
		The circulation and temperature changes during the SSW event induce variations in trace species. This process results in an increased CO$_2$ density due to upwelling in the polar MLT during the peak warming period. After the SSW event, the wave-induced downward winds bring the CO$_2$ poor air-mass into lower altitudes, resulting in reduced CO$_2$ density in the polar MLT, in the recovery phase. We have found an anti-correlation between CO$_2$ density and CO$_2$ IR cooling during and after the SSW event, which can be attributed to upwelling/downwelling-induced changes in temperature, O and CO$_2$ densities. The CO$_2$ IR cooling has been seen to decrease during the peak warming period, and increase after the SSW event. These cooling patterns during an SSW event are consistent with the changes in temperature and O density, which play a crucial role in the CO$_2$ IR cooling in the MLT. The variation in O density and mesospheric temperature causes changes in collisional excitation of the CO$_2$ into the higher vibrational state, thus resulting in CO$_2$ IR cooling alterations during the SSW event. The role of CO$_2$ change in the observed CO$_2$ IR cooling variation during the SSW peak warming and recovery periods is found to be insignificant. Therefore, we can conclude that the variation in the temperature is the major contributing factor in the observed variability of CO$_2$ IR cooling. This along with change in O density, due to upwelling/downwelling in the MLT, controls the observed variability of CO$_2$ IR cooling with CO$_2$ density, during the 2009 major SSW event. A comprehensive study of the trace species and cooling processes in the MLT and their variation during extreme dynamic events such as SSW events is essential for a better understanding of the energetics and dynamics of the MLT region.

				\section*{Data Availability Statement}
				The TIMED/SABER data can be downloaded from the following link: \url{http://saber.gats-inc.com/data.php} \cite{SABER2023}.
				The SD WACCM-X output data, as free-run history files, can be obtained from the Climate Data Gateway at NCAR at \url{https://www.earthsystemgrid.org/dataset/ucar.cgd.ccsm4.SD-WACCM-X_v2.1.html} \cite{WACCMX2023}, and ACE-FTS data has been obtained from \url{http://www.ace.uwaterloo.ca/data.php} \cite{ACEFTS}.

				\acknowledgments
				The authors acknowledge the TIMED-SABER instrument, algorithm, and data processing teams for providing free access to the SABER dataset and the SCISAT/ACE-FTS teams for providing the data used in this study. We sincerely thank the WACCM team for providing free-run modelled data of SD WACCM-X. The authors thank the DST-SERB and the Ministry of Education, Government of India, for the research assistantship.
				
				%
				
				
				%
				
				

				
				%
				%
				%
				%
				%

			\end{document}